# Combining Dynamic Analysis and Visualization to Explore the Distribution of Unit Test Suites

Amjed Tahir[1] and Stephen G. MacDonell[1,2]

[1]*Department of Information Science, University of Otago, Dunedin, New Zealand*
[2]*School of Computer and Mathematical Sciences, Auckland University of Technology, Auckland, New Zealand*
*amjed.tahir@otago.ac.nz, stephen.macdonell@otago.ac.nz*

## Abstract

*As software systems have grown in scale and complexity the test suites built alongside those systems have also become increasingly complex. Understanding key aspects of test suites, such as their coverage of production code, is important when maintaining or reengineering systems. This work investigates the distribution of unit tests in Open Source Software (OSS) systems through the visualization of data obtained from both dynamic and static analysis. Our long-term aim is to support developers in their understanding of test distribution and the relationship of tests to production code. We first obtain dynamic coupling information from five selected OSS systems and we then map the test and production code results. The mapping is shown in graphs that depict both the dependencies between classes and static test information. We analyze these graphs using Centrality metrics derived from graph theory and SNA. Our findings suggest that, for these five systems at least, unit test and dynamic coupling information 'do not match', in that unit tests do not appear to be distributed in line with the systems' dynamic coupling. We contend that, by mapping dynamic coupling data onto unit test information, and through the use of software metrics and visualization, we can locate central system classes and identify to which classes unit testing effort has (or has not) been dedicated.*

**Keywords:** test analysis; program comprehension; unit testing; visualization; dynamic metrics; dynamic analysis

## 1. INTRODUCTION AND MOTIVATION

Achieving improvements in software testability, alongside other quality and productivity goals, is crucial to contemporary software development. It is widely acknowledged that software systems are growing larger and becoming more complex [1], and yet the resources directed towards testing are not keeping pace [2]. This is because testing is known to be an expensive process, one that can consume upwards of 50% of the total time and cost needed for software development [3]. Although such figures are typically associated with waterfall-like processes where testing is treated as a 'phase', the centrality of testing is not just a phenomenon of plan-based development approaches: Agile software development methods such as eXtreme Programming (XP) also give testing significant attention [4]. The practice of Test-Driven Development (TDD), for example, requires that extensive test code be developed and maintained to ensure that the 'furthermost' components of the production code work correctly [5]. In these methods, in fact, unit tests are viewed as core, integral parts of the program [6]. In noting the importance of testing, Beck [7] recommended that developers spend between 25% and 50% of their time writing tests.

Irrespective of the development method adopted, then, testing is a high-cost activity. That said, effective testing is also a high- value activity. Unit tests – which exercise individual (or small groups of) software units (e.g., software classes) – provide a powerful mechanism for validating existing features when in the process of developing new functionality [6]. When well- designed, the use of unit tests is known to improve software quality from the early stages of development and to enable the detection of defects more effectively when compared to other verification strategies [8]. The ideal ratio of test code to production code (particularly in systems implemented with methods similar to TDD) is said to be 1:1; however, the typical ratio in OSS is estimated to be 2:3 [9] or less. This latter ratio suggests that unit tests are generally not available for all production classes in OSS systems. Thus, in a typical use profile, production classes that are perhaps heavily executed may not have any directly associated unit tests.

As noted above, contemporary software systems can be challenging to understand [10], given complex interactions between classes and objects. Moreover, some of those interactions only become evident during program execution. For instance, complex structural features, such as inheritance, polymorphism and dynamic binding, as well as the presence of dead code, are likely to appear only when code is run [11]. As such, their incidence and effects can be precisely assessed only by using dynamic measurements rather than traditional static measurements. Equally, understanding test code can be a challenge due to the fact that tests are not always well-structured [12].

This research analyzes test adequacy in several OSS systems by augmenting standard static analysis approaches with dynamic analysis and visualization techniques. In this research, we collect and combine dynamic coupling and unit test data in order to provide a more complete picture of unit tests' distribution (i.e., the distribution of the unit tests over system's classes), across five software systems. Novel visual representations are developed to present the dynamic information directly in relation to unit tests. The work thus contributes a novel view that combines dynamic coupling and unit test data in order to support greater understanding of unit tests' distribution in OSS. The field of software visualization offers promise in aiding engineers to better understand certain aspects of software behavior [13]. It has also been suggested that software metrics should be examined through appropriate visualizations, thus achieving improved understanding beyond the 'raw' numbers of the metrics alone [14]. In short, using visualizations to support program comprehension and the understanding of software artifacts (including test artifacts) appears to be effective and useful [4, 10, 15-17].

We contend that visualizing such data could be especially helpful during maintenance and reengineering tasks, as the visualization process elucidates the hierarchy of the production classes and the distribution of unit tests corresponding to the production classes. In addition, such visualizations would provide developers and testers with a high-level view of the dependencies within a system and the possible utilization of methods for future testing activities, i.e., what components are not being tested, and where testing effort should be focused.

The remainder of this paper is structured as follows. Section 2 presents background information on software testability and dynamic software metrics, respectively, followed by a summary of related work. The study design and our research objectives are presented in Section 3. Section 4 describes the data collection process and the OSS systems evaluated in this research. This is followed by the results and a discussion of those results in Section 5. Threats to validity are presented in Section 6. Finally, Section 6 concludes the study and presents thoughts on future work.

## 2. BACKGROUND AND RELATED WORK

Prior to describing our own empirical work in detail we now provide background information on the nature of software testability followed by a discussion of the importance of dynamic metrics in relation to the work reported here. We also review several works related to this study.

### A. Software Testability
It follows logically that improving the testability of software should enable developers to achieve higher quality outcomes for that software. However, defining, measuring and then improving testability present significant challenges in their own right. Like many non-functional properties of software, testability has been acknowledged as an elusive concept, and its measurement and evaluation have been considered to be inherently difficult [2]. Although several standards and individual studies have defined testability, they have done so in various ways, reflecting the fact that they were motivated by different purposes. Thus, testability has been defined based on test effort, test coverage or the ability to examine if requirements are met. For example, the IEEE standard defines testability as "the degree to which a system or component facilitates the establishment of test criteria and the performance of tests to determine whether those criteria have been met". Another definition is "the degree to which a system can be unit tested and system tested" [18]. The relevant ISO standard [19] defines testability as "attributes of software that bear on the effort needed to validate the software product". Thus, the IEEE definitions consider software testability from a test criteria point of view. The ISO definition, in contrast, considers testability based on the effort needed to test a software product.

Software classes with low levels of testability may be less trustworthy, even after successful testing [20]. Classes with poor testability are also more expensive to repair when problems are detected late in the development process. In contrast, classes with good testability can dramatically increase the overall quality of the software, and reduce the cost of testing [21]. Some researchers relate software testability and test efficiency to the effort and cost of conducting those tests [2, 21]. Testability has also been related to internal characteristics of software systems, including various attributes of software design and code [22-24].

### B. Dynamic Metrics
Dynamic metrics, which are used to capture the dynamic behavior of a software system, have been found to be directly related to a range of software quality attributes, including complexity, maintainability and testability [25, 26]. Their use has gained traction given that traditional static software metrics may not be sufficient for characterizing and predicting the quality of OO systems [27-29]. Dynamic metrics are computed based on data collected during program execution (i.e., at runtime) and are most frequently obtained from the execution traces of the code (although in some cases simulation can be used instead of the actual execution), and therefore they can directly reflect the quality attributes of a system in operation [30]. Our recent survey on dynamic measurement research shows that dynamic metrics are attracting growing attention from researchers, mainly because of the inherent advantages of this class of metrics over their static counterparts [30].

That is not to say that static metrics have no value, and this value may be elevated further if they are combined with data collected through dynamic analysis [31]. The two forms are complementary; they should therefore be used alongside one another to build strong affordances about the software under investigation. We follow this thinking and combine static test information with code run-time properties collected during software execution. Specifically, we collect a dynamic coupling metric that is based on runtime method invocations. Method invocation information of system classes is captured and then visualized using dependency graphs. This is explained in more detail in Section 3 (C. Metrics Definition).

**C. Related Work**
Extensive effort in both software engineering research and practice has been directed to supporting the understanding and maintenance of software artifacts. Of particular relevance here is work that has used static and/or dynamic analysis techniques. While the two approaches were initially used separately, combining static and dynamic analysis techniques has been of growing interest. Many works (including [10, 32]) have proposed and assessed various methods, techniques and tools that use data obtained from both static and dynamic analysis to support program understanding. Visualization, in particular, has been used in several previous works for the purpose of supporting developer understanding of different aspects of production code [10, 15, 16]. Test understanding has been the focus of several works that sought to explicate the relationship between production and test code, while other works have considered the structure of unit tests and test suites. Visualization of test information as a means of supporting developer understanding has also been considered in prior work [4, 17]. We now briefly discuss the most relevant of these prior related studies.

Cornelissen et al. [4], based on information obtained through dynamic analysis (though generated via simulation), used UML sequence diagrams to visualize test cases to gain knowledge about the structure of software in order to support program understanding. They asserted that such visualizations could be beneficial in program understanding and for documentation purposes. Visualization of test code dependencies was used by van Rompaey and Demeyer [17] to localize unit tests and to investigate the relationship between test and production code. Their focus was on both the composition of and dependency between test and production units as well as among the unit tests themselves. The dependency information was obtained from static properties of the system. Although a number of coupling and cohesion indicators were recorded, they were used only to identify the dependency between classes and their associated unit tests. However, the authors recommended that size and complexity information of the various software components should also be considered to provide a more detailed and comprehensive assessment of the proposed visualization approach. In similar work, Zaidman et al. [33] used visualization to investigate the co-evolution between software production code and test suites. Their study focused on mining software history information from repositories in order to detect testing information from different versions of software projects. The authors also observed a signification correlation between test effort (i.e., test-writing activity) and test coverage levels in different releases. The work also proposed three different visualization views that could be used to study how test code co-evolves over time between different releases.

Bruntink and van Deursen [23] used several static OO complexity metrics to measure class-level testability in order to inform the planning and management of subsequent testing activities. Their empirical study found a strong correlation between a number of static class-level measures and their defined testability measures. In following these findings we also identified some significant relationships between dynamic software properties (represented in terms of dynamic coupling and execution frequency) and the same class-level testability measures suggested by Bruntink and van Deursen [23] in a recent study of our own [24]. Hauptmann et al. [12] used a clone detection technique to identify and locate tests in order to support better understanding of these tests. The technique was applied to 4000 tests across seven industrial systems. In general, clone detection was found to provide useful information for targeting test automation effort. The findings also revealed that significant numbers of clones exist in all examined "manually written" tests.

In following the studies just described we build on these works (specifically those works that used visualization and/or dynamic analysis to support comprehension and understanding such as [4, 17]) in terms of the metrics used and visualization support provided. The key elements of the work conducted and reported in this paper include the following:

- The metric data collected in this work include a dynamic coupling measure that represents the run-time dependencies between classes/objects;

- The visualization provided in this work combines both the dynamic coupling information collected from the production code with static test data collected from the associated unit tests;

- Graph metrics are used to characterize the visualizations and so provide additional insights.

The design of the study is now presented in detail.

# 3. STUDY DESIGN

As noted in Section I, understanding test code is an important task in software development, particularly in relation to the activities of maintenance, reverse engineering and refactoring. In the object-oriented paradigm, production code and test code are similar in nature (i.e., written in a similar manner); thus, analyzing and understanding them requires similar skills and methods. In this section, we state our research objectives, we describe our data collection methods and we specify the OSS systems analyzed.

**A. Objectives**
The main objectives of this work are to:

- Examine the utility of combining dynamic and static information to expose test distribution.

- Represent test distribution information in a visualization that combines static and dynamic analysis data.

- Demonstrate the application of the visualization on sample OSS systems, including systems of different size.

In achieving the above objectives this research will enable us to assess whether dynamic information, here represented by dynamic coupling, might be useful when added to unit test information to represent the distribution of unit tests in a sample of software systems. A key aim of this work is to determine whether production classes and unit tests are evenly distributed; that is, do all highly and/or tightly coupled classes have dedicated unit tests and test classes? A secondary aim is to develop a new visualization that combines dynamic information associated with production code and test information, with a view to supporting better understanding of the distribution of test suites in software systems.

### B. Contributions
The findings of this work contribute to the general body of knowledge on software understanding (and more specifically, test understanding) by visualizing a new combination of static and dynamic information that could aid the test understanding process. The methods developed in this study should provide developers with knowledge of the unit testing distribution and activities in OSS systems.

One possible use of the proposed visualization is when maintenance and reengineering activities are planned. The visualization should enable engineers (and in particular, maintainers and reengineers) to explore the distribution of unit tests in relation to the dynamic behavior of the software before conducting their work. It should also benefit program understanding by providing a visual representation of the dependencies based on the actual use of the properties of the system. Newcomers to a project could also use these visualizations to understand which aspects have been directly covered with unit tests in relation to their dynamic dependencies view [17]. The proposed visualization could also be beneficial for Agile-like methods, in which tests (and in particular unit tests) serve as documentation [4].

### C. Metrics Definition
Coupling has long been shown to have a direct impact on the quality of software, primarily through its relationship to the characteristics of complexity and maintainability. Two classes/objects are said to be coupled if at least one of them acts upon the other [34]. All other things being equal, the greater the coupling level of an artifact, the greater the complexity, and the harder it is to maintain that artifact [35]. As stated above, previous research also suggests that coupling has a direct impact on class-level testability [2, 24]. For many years coupling has been measured statically, based on various structural properties of software [36]. While undoubtedly useful, this naturally does not account for coupling between objects at runtime. By also considering this information, we should be able to obtain a more complete picture of the 'true' testability of a system.

In recent years dynamic coupling has received increasing research attention [30]. In this study we measure dynamic coupling based on runtime method invocations/calls. Assuming there are two classes, class A and class B, A and B are said to be coupled if a method from class A (*caller*) invokes a method from class B (*callee*). We use the Dynamic Coupling Between Objects (DCBO) metric to represent these data. As the name implies, DCBO is the dynamic form of the well-known Coupling Between Objects (CBO) metric [34]. In this research, we use an automatic code instrumentation method represented in Aspect Oriented Programming (AOP) to collect this dynamic information, in line with several previous works (e.g., [35-37]). Metrics data are collected using the AspectJ framework - a well-established Java implementation of AOP. For any class, the DCBO metric computes the total number of classes that are invoked by that class during execution.

## 4. DATA COLLECTION AND OSS SYSTEMS' SELECTION

Through the use of tools and plugins we measure the size of the OSS systems selected for this study using a range of static metrics. The Kilo Lines of Code (KLOC) and Number of Classes (NOC) metrics were collected by the *CodePro Analytix*[1] static analysis tool and were subsequently verified using the *Eclipse Metrics Plugin*[2] (by comparing the values of the metric data collected by the two tools and resolving the few discrepancies). The Number of Test Cases (NTC) metric was collected through the JUnit framework and its results were verified manually by the first author. The author checked the naming of the unit test and, if needed, the calls made from unit tests. Dependency graphs were generated using NodeXL. Test coverage data were collected using both the CodeCover and Emma coverage tools.

To identify unit tests and associate them with their corresponding production classes we used two different established traceability techniques [38]. First, we used the Naming Convention technique, which reflects the widely suggested practice (for instance, in the JUnit documentation) that a unit test should be named after the corresponding class(es) that it tests, by adding "Test" to the original class name. Second, we used a Static Call Graph technique, which inspects method invocations in the test case. The latter process was carried out manually by the first author. The effectiveness of the Naming Convention technique is reliant on developers' efforts in conforming to the recommended coding standard, whereas the Static Call Graph approach reveals direct references to production classes in the unit tests. It is important to note here that we consider core system code only. That is, only production classes that are developed as a part of the system are assessed. Additional classes (including those in jar files) are excluded from the measurement process.

We identified five OSS systems to be used in this study, via the following selection criteria: all software systems must: 1) be fully open source, 2) have unit tests available, 3) be written in Java, and 4) cover a range of systems sizes. For the latter criterion we used a classification motivated by the prior work of Zhao and Elbaum [39], but with changes in its structure in order to meet the growing scale of OSS systems. We therefore categorized application sizes into bands based on the number of KLOC: tiny (fewer than 1 KLOC), small (1 up to 10 KLOC), medium (10 up to 100 KLOC), large (100 up to 1000 KLOC), and extra-large (more than 1000 KLOC). Our initial aim was to have at

---

TABLE I. GENERAL CHARACTERISTICS OF THE SELECTED OSS SYSTEMS

| System | Version | URL | Size | KLOC | NOC | # JUnit classes | NTC | Test KLOC | Statement Coverage | Class Coverage |
|--------|---------|-----|------|------|-----|-----------------|-----|-----------|--------------------|----------------|
| FindBugs | 2.0.3 | http://findbugs.sourceforge.net | Large | 114.481 | 1,245 | 42 | 191 | 2.522 | 13.9% | 26.5% |
| JabRef | 2.9.2 | http://jabref.sourceforge.net | Medium | 84.717 | 616 | 55 | 237 | 5.392 | 29.6% | 46.7% |
| Dependency Finder | 1.2.1 beta4 | http://depfind.sourceforge.net | Medium | 26.231 | 416 | 258 | 2,003 | 32.095 | 59.8% | 59.5% |
| MOEA | 1.17 | http://www.moeaframework.org | Medium | 24.307 | 438 | 280 | 1,163 | 16.694 | 77.2% | 86.5% |
| JDepend | 2.9 | http://clarkware.com/software/JDepend | Small | 2.460 | 29 | 18 | 108 | 1.100 | 41.3% | 41.3% |

TABLE II. BRIEF DESCRIPTIONS OF THE SELECTED SYSTEMS

| System | Description |
|--------|-------------|
| FindBugs | Widely used static code analyzer that analyses Java bytecode to find and detect a wide range of pre-defined bugs and defects. The tool includes more than 200 bug patterns. |
| JabRef | Bibliography tool that provides GUI-based reference management support for BibTeX files - the standard LaTeX bibliography format. |
| Dependency Finder | An analyzer that extracts dependencies and dependency graphs of complied Java code and mines some other useful dependency information. The tool also provides basic OO quality metric assessment of source code. |
| MOEA | Java-based framework that supports development and experimentation of multi-objective evolutionary and optimization algorithms. The tool is intended to provide fast and reliable implementations of several state-of-the-art multi-objective algorithms. |
| JDepend | Light analysis tool that evaluates Java packages using several OO quality metrics. The tool provides an automated way to measure the quality of software design. |

least one project fit into each of the small, medium and large size categories, as considering systems of different size should enable us to test our approach at different scales. The systems selected for our experiment are: FindBugs, JabRef, Dependency Finder, MOEA and JDepend. Table I shows some of the general characteristics of the selected systems, including size information in KLOC and NOC for the production code, and NTC and Test KLOC for the test code. As shown, out of the five selected systems, one is large, three are medium and one is a small-sized system. Table I also shows test coverage information for the five systems. Statement coverage levels vary from 13.9% in the case of FindBugs to 77.2% for MOEA. MOEA also has the highest class coverage at 86.5%, where the lowest class coverage value is recorded for FindBugs at 26.5%.

In order to arrive at dynamic metrics values that are associated with typical, genuine use of a system the selected execution scenarios must be representative of such use. Our goal is to mimic 'actual' system behavior, as this will enhance the utility of our results. Execution scenarios are therefore designed to use the key system features, based on the available documentation and user manuals for the selected systems, as well as our prior knowledge of these systems. We present here the details of the specific execution scenario developed for each system. A brief description of all five selected systems is provided in Table II. Graph data packages for all five systems are available for replication purposes[3].

### FindBugs

We used FindBugs' main GUI tool to analyze jar and source code files of three other Java OSS systems, two of large size (i.e., FindBugs itself and Apache JMeter) and one of medium size (i.e., Dependency Finder). During the execution, we activated cloud-based storage by loading the tool's external cloud plugin. Finally, all analysis reports were then exported for all three systems in various formats.

### JabRef

The tool is used to generate and store a list of references from an original research report. We included all reference types supported by the tool (e.g., journal articles, conference proceedings, reports, standards). Reports were then extracted using all available formats (including XML, SQL and CSV). References were managed using all the provided features. All additional plugins provided at the tool's website were added and used during this execution.

### Dependency Finder

This scenario involves using the tool to analyze the source code of three large systems: FindBugs, Apache JMeter, and Apache Ant) and one medium-sized system (Colossus). We computed dependencies, dependency graphs and OO metrics at all layers (i.e., packages, classes, features). Analysis reports were extracted and saved individually in all possible formats.

### MOEA

MOEA has a GUI diagnostic tool that provides access to a set of algorithms, test problems and search operators supporting multi-objective optimization. We used this diagnostic tool to apply those algorithms on all the predefined test problems. We applied each of the algorithms at least once to each problem. We displayed metrics and performance indicators for all results provided by those different problems and algorithms. Statistical results of these multiple runs were displayed and saved at the end of the run.

### JDepend

We designed a small GUI to provide access to all of the quality assessment and reporting functionalities of JDepend (Note: this additional code was excluded from the measurement collection and analysis.) The tool was then show only a snapshot of the dependency graph. Given that all visualization graphs should be presented in a complete used to load and analyze four OSS systems, three of medium size (i.e., Dependency Finder, JabRef, and barcode4j) and one of large size (i.e., FindBugs). We used all three user interfaces provided by the tool (namely: swing-graphical, textual and XML) during this execution.

---

[3] http://goo.gl/nuGZ4u

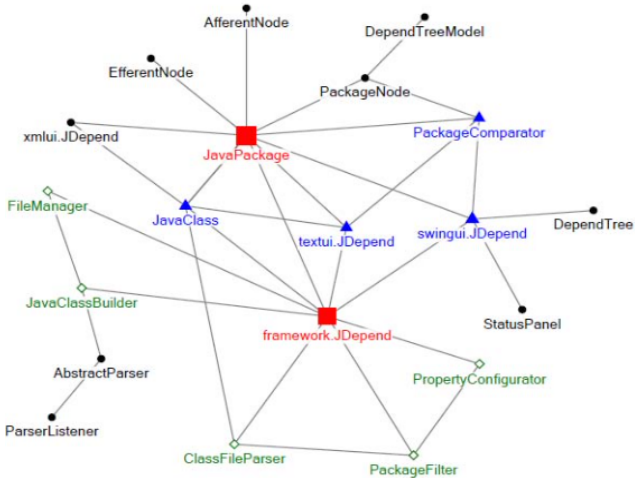

Fig. 1. JDepend full Dependency Graph

TABLE III. DEPENDENCY GRAPH NODE SYMBOLS

| Symbol | Description |
|---|---|
| ■ | Tightly coupled class, with at least 1 dedicated unit test |
| ▲ | Tightly coupled class, with no dedicated unit test |
| ◇ | Loosely coupled class, with at least 1 unit test |
| ● | A production class with no dedicated unit test |

TABLE IV. CENTRALITY METRICS FOR JDEPEND

| Class | Degree Centrality | Betweenness Centrality | Unit test? |
|---|---|---|---|
| framework.JDepend | 9 | 92.8 | Yes |
| JavaPackage | 9 | 72.6 | Yes |
| swingui.JDepend | 5 | 37.5 | No |
| JavaClass | 5 | 8.6 | No |
| PackageComparator | 4 | 5.0 | No |
| textui.JDepend | 4 | 3.2 | No |
| JavaClassBuilder | 3 | 34.0 | Yes |
| PackageNode | 3 | 18.0 | No |
| ClassFileParser | 3 | 0.8 | Yes |
| PackageFilter | 3 | 0.5 | Yes |
| FileManager | 2 | 0.5 | Yes |
| AbstractParser | 2 | 18.0 | No |
| PropertyConfigurator | 2 | 0 | Yes |
| xmlui.JDepend | 2 | 0 | No |
| ParserListener | 1 | 0 | No |
| AfferentNode | 1 | 0 | No |
| DependTree | 1 | 0 | No |
| DependTreeModel | 1 | 0 | No |
| EfferentNode | 1 | 0 | No |
| StatusPanel | 1 | 0 | No |

## 5. RESULTS AND DISCUSSION

This section presents the results, analysis and discussion of our empirical investigation of the proposed measurement and visualization approach. A dependency graph is used to visually depict the dependencies between classes in each of the five systems. Dependencies, shown as *undirected* edges, represent method invocations between classes, shown as nodes. An undirected edge between nodes A and B means that the two nodes are coupled. That is, a dependency between classes A and B represents at least one invocation from a method in class A to a method in class B, and/or vice versa. A description of the dependency graph node symbols is provided in Table III. For tightly coupled classes the size of the vertex represents the relative degree of coupling.

To compare node (class) dependency, and to generally quantify the level of association that a node has with other nodes in the graph, we measure their *Centrality*. Centrality is a well-known concept in graph theory and has been used increasingly in recent times in Social Network Analysis (SNA). For each node on the graph we collect two metrics: Degree Centrality and Betweenness Centrality. Degree Centrality measures the total number of links (connections) for a node. This metric directly reflects the dynamic coupling information, which is obtained from the DCBO metric, demonstrating messages sent or received by a class (also known as Import and Export Coupling). Betweenness Centrality, on the other hand, is a measure of the number of times a node acts as a bridge between two other nodes in the graph.

Figures 1 to 5 show dependency graphs for all five systems. Due to space limitations, however, we show full dependency graphs only for the JDepend and Dependency Finder systems, whereas for the other three systems we form and seen in clear coloring we provide full high-resolution versions elsewhere[4].

As shown in Table IV and visually in Figure 1, the *framework.JDepend* and *JavaPackage* classes of the JDepend system are shown to have the highest levels of (Degree and Betweenness) Centrality. Both classes are also directly tested through dedicated unit tests. Other classes, including *swingui.JDepend* and *JavaClass*, have high levels of Degree Centrality (both have the second-highest value) but have no associated unit tests. We also note that the *FileManager* and *PropertyConfigurator* classes have dedicated unit tests associated with them even though they are not shown to be central to the system's operation (i.e., their Centrality levels are low, especially in terms of Betweenness Centrality).

A comparison of the different levels of Centrality for tested classes across our five selected OSS systems is presented in Table V. In this table we compare results from a proportion of the classes from the 'top' and 'bottom' of their ranked lists. That is, for each system, we rank the Centrality data values and then divide the data into four groups based on three quartile data points. The 1st (Q1 - the lower) and the 3rd (Q3 - the upper) quartiles split off the bottom and top 25% of the data points in terms of centrality, respectively, whereas the 2nd quartile (Q2 - the median) reflects the middle 50% of the data. Those classes above the Q3 threshold are relatively highly coupled, and those below the Q1 threshold are coupled loosely.

TABLE V. LEVELS OF CENTRALITY IN THE EXAMINED SYSTEMS

| System | Number and proportion of tested classes above Q3 for Degree Centrality | Number and proportion of tested classes above Q3 for Betweenness Centrality | Number and proportion of tested classes below Q1 for Degree Centrality | Number and proportion of tested classes below Q1 for Betweenness Centrality |
|---|---|---|---|---|
| **FindBugs** | 11 | 4 | 1 | 2 |
| | 7% | 2% | 1% | 1% |
| **JabRef** | 9 | 11 | 9 | 4 |
| | 13% | 15% | 13% | 6% |
| **Dependency Finder** | 31 | 24 | 3 | 19 |
| | 69% | 53% | 7% | 42% |
| **MOEA** | 25 | 21 | 17 | 20 |
| | 66% | 55% | 45% | 53% |
| **JDepend** | 2 | 3 | 0 | 2 |
| | 40% | 60% | 0% | 40% |

TABLE VI. CENTRALITY METRICS MANN-WHITNEY U TEST RESULTS WITH EFFECT SIZE

| Metrics | | FindBugs | JabRef | Dependency Finder | MOEA | JDepend |
|---|---|---|---|---|---|---|
| Betweenness Centrality | $\alpha$ | **0.00** | 0.50 | **0.00** | 0.08 | 0.11 |
| | ez | 0.12 | 0.04 | 0.27 | 0.14 | 0.36 |
| Degree Centrality | $\alpha$ | **0.01** | 0.16 | **0.03** | 0.45 | 0.26 |
| | ez | 0.10 | 0.08 | 0.16 | 0.06 | 0.25 |

For Dependency Finder (Figure 2) Centrality values for classes above the Q3 threshold (being 46 classes, with some exceeding the threshold for both Centrality measures). We found that almost half of the classes had no associated unit tests. For example, the dependency.Printer class has a Degree Centrality value of 53 (the second highest in the system) and its Betweenness Centrality is ranked 5th highest in the system, yet it has no associated unit tests. This class is considered to be central to the system based on its dynamic coupling values. In contrast, we observed other classes with very low levels of Centrality but with dedicated unit tests. For example, the RegularExpressionParser and PrinterBuffer classes both have devoted unit tests even though they have the lowest Centrality values, with only a value of one for Degree Centrality and zero for Betweenness Centrality. This latter result indicates that these classes do not appear to be central to the system's operation in terms of their dynamic coupling.

A generally similar pattern of unit tests' distribution is evident in all five systems examined. Figures 3 through 5 show a snapshot from the FindBugs, JabRef and MOEA dependency graphs, respectively. In considering the Q3 classes by Centrality in JabRef (71 classes) we found only nine (among 71) to have devoted unit tests. Similarly, only eleven classes with the highest Betweenness Centrality measure were found to have dedicated unit tests.

In regard to the MOEA system, unit tests are present for 25 (66%) of the classes above the Q3 Degree Centrality threshold and for 21 (55%) classes above the Q3 value for Betweenness Centrality. However, MOEA also has the highest proportions of tested classes below the Q1 Centrality measure thresholds, with 45% and 53% of these classes having unit tests. This may be a reflection of the generally high levels of test coverage in MOEA (see Table

II). The lowest percentages of tested classes above Q3 for both Degree and Betweenness Centrality are evident for FindBugs (although it is also the largest of the five systems examined). It has 11 (~7%) classes with associated unit tests among the 164 classes in Q3, and only 1 tested class (< 1%) in Q1 for the Degree Centrality classes. For Betweenness Centrality, there are 4 tested classes in Q3 and 2 classes in Q1.

To provide a more comprehensive assessment of the relationship between the Centrality metrics values and the availability of unit tests for production classes we used the nonparametric (two-tailed) *Mann-Whitney U* test (as the data come from non-normal distributions). We investigate the following hypothesis: *"there is a significant difference between centrality metrics values of production classes with associated unit tests and those without associated unit tests"*. We also measure the effect size (ez) using the following nonparametric formula (1)[40]:

$$ez = Z / \sqrt{N} \qquad (1)$$

where N is the number of observations, Z is the z-value (also known as the standard score).

We classify the effect size using Cohen's classification: *small effect size when $0 < ez < 0.3$, medium when $0.3 \leq ez < 0.5$ and large when $ez \geq 0.5$.*

We report the values of our Mann-Whitney U test and effect size for TLOC and NTC in Table VI. Significant p-values ($\alpha$) are shown in **bold**. We found significant $\alpha$ in only two of the five examined systems (i.e., FindBugs and Dependency Finder). Even through the p-values are significant in these systems, the effect size values are small for both Centrality metrics. The other three systems did not show any significant values. We therefore reject the hypothesis and conclude that there is no significant difference between Centrality metrics values of production classes with associated unit tests and those without associated unit tests.

Several observations can be made based on the results just presented. The main observation, enabled by the visual representations of the dependency graphs and the centrality measurements, is that there is no statistically significant relationship between dynamic coupling (and centrality metrics) and unit test coverage – unit tests do not appear to be distributed in line with the systems' dynamic coupling. In the five OSS systems examined, it is evident that many classes (i.e., more than 40% of the classes as shown in three of the five examined systems) that are loosely coupled and have few connections have received testing attention and effort (i.e., they have dedicated unit tests). Loosely coupled classes (shown at the outside of the graphs) have fewer connections and so are not intensively accessed by other classes. On the other hand, high proportions of classes in each system (up to 69% of the classes as shown in Dependency Finder) that are tightly coupled (i.e., highly linked/accessed by other classes) have no dedicated unit tests.

Of particular note is that this distribution pattern is present in all five systems, regardless of their test coverage levels. However, the specific numbers of tested and untested classes varies from one system to another. From Table IV it is evident that the proportion of unit tests in relation to coupling Centrality levels is different in all five systems. This suggests that, even in mature OSS systems such as these, the dispersed nature of contributions to the project may mean that test distribution can be uneven, and provision of tests is reliant on the attention of the individual developers.

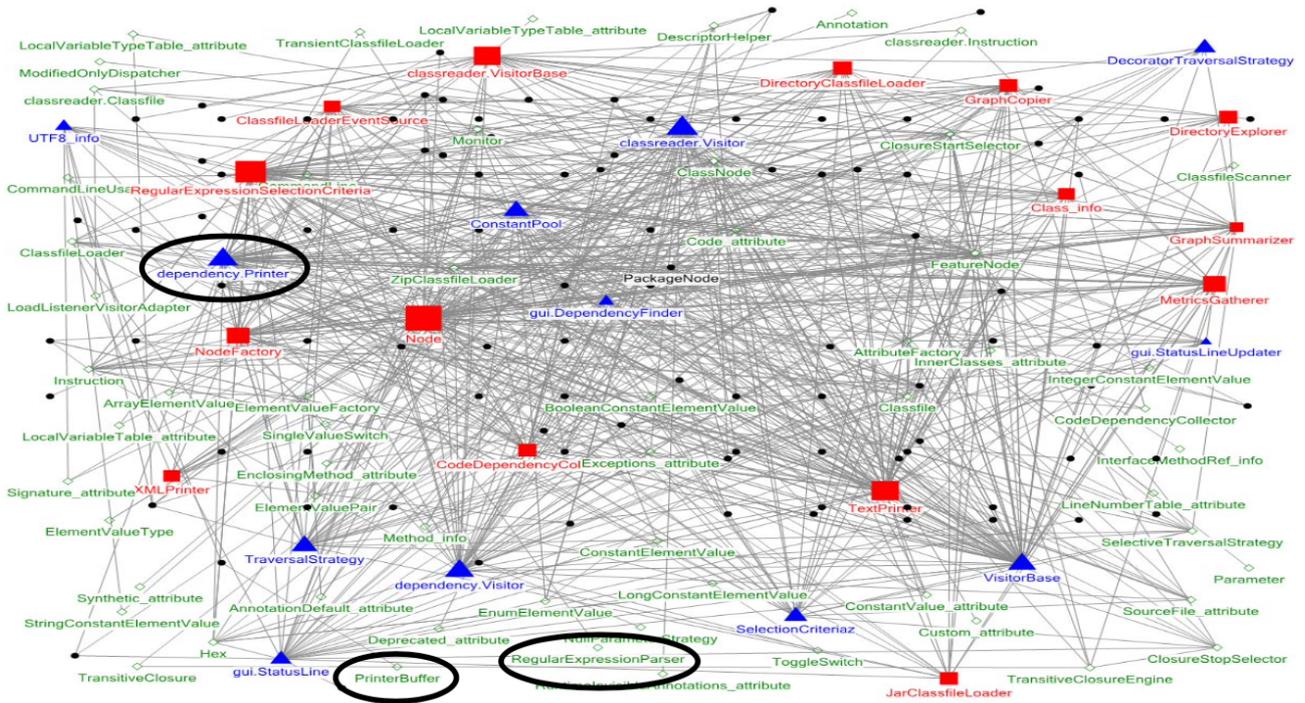

Fig. 2. Dependency Finder full dependency graph.

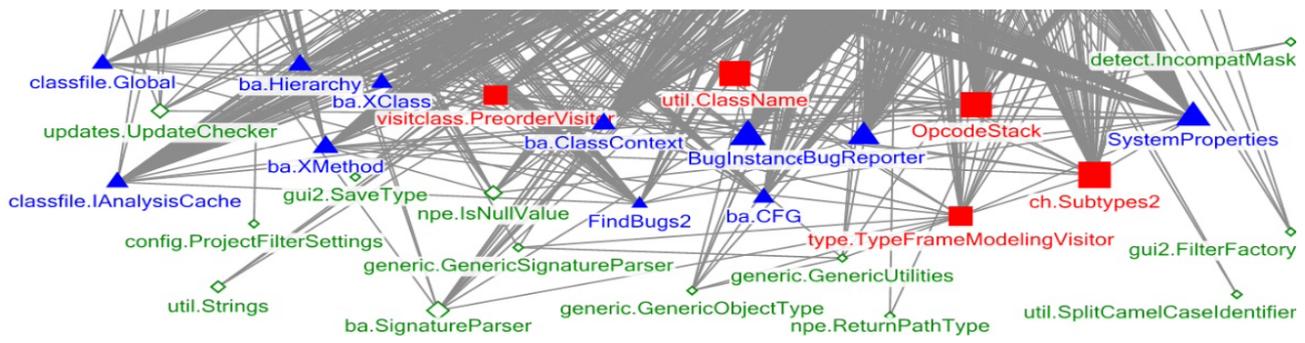

Fig. 3. FindBugs dependency graph snapshot.

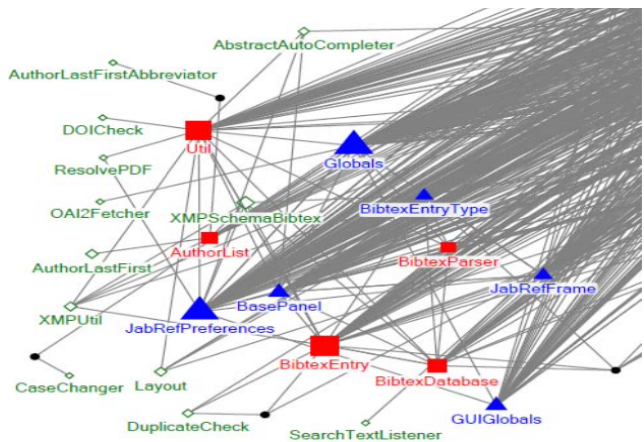

Fig. 4. JabRef dependency graph snapshot.

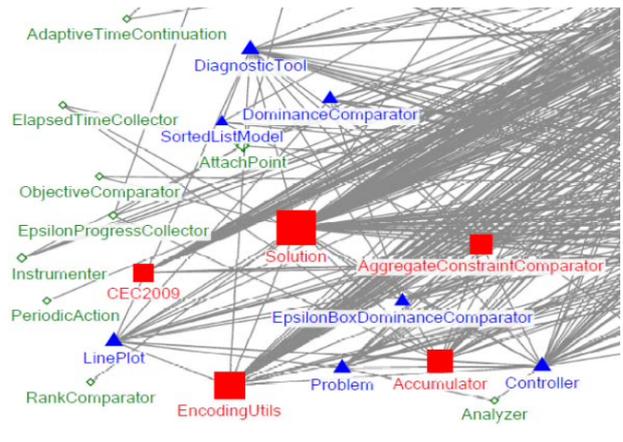

Fig. 5. MOEA dependency graph snapshot.

The results presented here suggest that the distribution of unit tests may require more attention from engineers/testers but also from those managing software development. We contend that the suggested visualization will help in focusing and optimizing testing effort by allowing engineers to identify and initially target central system classes and to dedicate relatively less effort to those non-central classes. We also suggest that the centrality metrics themselves could be helpful in providing quantitative support for the visualizations of the dependency graphs. The two Centrality metrics provided us with a comprehensive insight into the levels of dependency between system classes.

## 6. THREATS TO VALIDITY

We acknowledge a number of threats to the validity of our study. One of the possible threats is the selection of the execution scenarios. All of the execution scenarios used here were designed to mimic as closely as possible 'actual' system behavior, based on the available system documentation and, in particular, indications of each system's key features. We acknowledge, however, that the selected scenarios might not be fully representative of the typical uses of the systems. Analyzing data collected based on different scenarios might give different results. This is a very common threat in most dynamic analysis research. However, we worked to mitigate this threat by carefully considering user manuals and other documentation of each of the examined systems. Most listed features were visited (at least once) during the execution. We will examine other scenarios in the future and compare the results from these different scenarios.

Another possible threat to validity is the limited number and form of the OSS systems investigated. Results discussed here are derived from the analysis of five OSS systems. The consideration of a larger number of systems, perhaps including closed-source systems as well as larger systems, would enable further evaluation of our results.

Varied availability of testing information could be another validity threat in our study. We used the available test information for the five systems in our analysis, and as such we did not have access to any information about the testing strategy employed. Test strategy and criteria information could be informative if combined with the test metrics, given that test criteria can inform testing decisions, and the number of test cases designed is highly influenced by the selected test strategy. Moreover, a more comprehensive picture of the analysis could be gained by also considering indirect tests.

Finally, we did not direct any attention to test quality – our interest at this stage is in the existence or otherwise of unit tests for system classes. An analysis approach that considers both the quantity and quality of the tests developed would seem likely to be optimal, however, and will be the subject of our future research.

## 7. CONCLUSIONS AND FUTURE RESEARCH DIRECTIONS

In this paper we introduced a visualization that combines dynamic information obtained from production code with static test information to depict the distribution of unit tests in OSS systems. Five such systems of different sizes were selected for examination in this study. We extracted the DCBO metric to measure dynamic coupling at execution time through AOP. We then collected basic unit test information using automated tools as well as manual traceability and verification methods. We generated dependency graphs to show dependencies between classes using the collected dynamic coupling information. Test information was then added to the dependency graphs to show how unit tests were distributed in comparison to the dynamic coupling information. The goal of this visualization is to assist reengineers and maintainers – and their managers – to observe and understand the distribution of unit tests in a software system based on a dynamic view of that system. The visualization is further supported by the use of graph Centrality metrics that provide insight into the production classes and unit tests distribution.

Based on the five OSS systems studied, we observe that unit test and dynamic coupling information 'do not match' in that there is no significant relationship between dynamic coupling and centrality metrics and unit test coverage. In other words, unit tests do not appear to be distributed in line with the systems' dynamic coupling. Many of the tightly coupled classes do not come with any associated direct unit tests, whereas other loosely coupled classes, which do not look to be central, appear to have received direct testing effort.

Visualization of the combined static test code data and dynamic coupling measurement data can provide a detailed view of how unit tests are actually distributed in relation to the coupling level of each class in the system. The suggested visualization and its associated Centrality metrics may help developers and managers to focus and optimize their test effort through the initial targeting of central system classes. Furthermore, data gathered from dynamic coupling measurement provides a comprehensive view of the dependencies of the system in relation to test information – a view that can be obtained only during software execution.

This work is still at an early stage and further validation of the approach is needed. Future work should also investigate the cause of this uneven distribution. It is important to examine this visualization approach with real software developers to evaluate the usefulness of the proposed approach in terms of improving program comprehension and understanding processes. One possible way to investigate this work would be through the use of a controlled user study/experiment with real software developers and maintainers. This study can also be extended by including testing strategy and indirect testing information into the mapping, to provide a more comprehensive view of testing activities and their distribution.

### ACKNOWLEDGMENT

We would like to thank Ewan Tempero and Abbas Tahir for their comments and constructive feedback on earlier versions of this work.